\documentclass[11pt,a4paper]{article} 
\usepackage{jcappub}
\usepackage{times}

\newcommand{\be}{\begin{equation}}
\newcommand{\ee}{\end{equation}}
\newcommand{\bea}{\begin{eqnarray}}
\newcommand{\eea}{\end{eqnarray}}
\newcommand{\bc}{\begin{center}}
\newcommand{\ec}{\end{center}}

\newcommand{\msun}{M_{\odot}}

\begin{document}
\title{Neutrino Halos in Clusters of Galaxies and their Weak Lensing Signature} 
\author[a]{Francisco Villaescusa-Navarro,} \author[b,c]{Jordi Miralda-Escud\'e,} \author[a]{Carlos Pe\~na-Garay,} \author[d]{Vicent Quilis}
\affiliation[a]{IFIC, Universidad de Valencia-CSIC, Valencia}
\affiliation[b]{Instituci\'o Catalana de Recerca i Estudis Avan\c cats, Barcelona}
\affiliation[c]{Institut de Ci\`encies del Cosmos (UB-IEEC), Barcelona}
\affiliation[d]{Departamento de Astronomia y Astrofisica, Universidad de Valencia, Valencia}

\emailAdd{villa@ific.uv.es}
\emailAdd{miralda@icc.ub.es}
\emailAdd{penya@ific.uv.es}
\emailAdd{vicent.quilis@uv.es}

\abstract{We study whether non-linear gravitational effects of relic neutrinos on the development of clustering and
large-scale structure may be observable by weak gravitational lensing. We compute the
density profile of relic massive neutrinos in a spherical model of a cluster of
galaxies, for several
neutrino mass schemes and cluster masses. Relic neutrinos add a small perturbation to
the mass profile, making it more extended in the outer parts. In principle,
this non-linear neutrino perturbation is detectable in an
all-sky weak lensing survey such as EUCLID by averaging the shear profile
of a large fraction of the visible massive clusters in the universe, or
from its signature in the general weak lensing
power spectrum or its cross-spectrum with galaxies. However, correctly modeling
the distribution of mass in baryons and cold dark matter and suppressing any
systematic errors to the accuracy required for detecting this neutrino perturbation is severely challenging.}

\maketitle

\section{Introduction}

The discovery of neutrino flavour conversion of solar, atmospheric,
reactor and accelerator neutrinos implies that at least two of the three
light neutrinos are massive. The sum of the neutrino masses is still
unknown. It is constrained from above ($\sim$ eV) by tritium beta decay
end point data and by cosmological data, and from below (0.05 eV) by
neutrino oscillation data. 
The neutrino mass squared differences are precisely measured by reactor and accelerator experiments, \cite{Concha}
\begin{eqnarray}
\Delta m_{21}^2=(7.6 \pm 0.2) \times 10^{-5}eV^2 ~,\\
\nonumber\\
\Delta m_{31}^2=(2.4  \pm 0.1) \times 10^{-3}eV^2 ~.
\end{eqnarray}
However, the neutrino mass hierarchy, or whether the two neutrinos with
the smallest mass difference are heavier or lighter than the other
one, is still unknown. Recent forecasts of galaxy clustering have included the neutrino mass ordering in addition to the total neutrino mass among the free 
model parameters that are considered, and show that future surveys
should reach the sensitivity required to explore most of the allowed
range of the total neutrino mass and to determine the neutrino
hierarchy \cite{Jimenez}.

Neutrino masses are usually included in the list of parameters of the standard model of cosmology in the linear regime, but this has rarely
been done in the nonlinear case. Massive neutrinos suppress the small scale matter power spectrum due to their large thermal velocities, making
the shape of the total mass power spectrum a potential probe to neutrino masses.
On scales much smaller than the free-streaming distance of neutrinos, the
relative suppression is \cite{Pastor},
\begin{equation}
\bigg|\frac{\Delta P(k)}{P(k)}\bigg|\simeq 10 \frac{\Omega_\nu}{\Omega_m},
\end{equation}
where $\Omega_{\nu}h^2 = (\Sigma_i m_i)/ (93.14 \hbox{ eV})$.

  Apart from this linear effect, massive neutrinos are also expected to
cluster around gravitationally collapsed dark matter haloes as their streaming
velocities are reduced and become comparable to the velocity dispersions
of the halos,
thereby modifying the dark matter halo total mass density profile.
Previous work has studied this non-linear neutrino clustering
\cite{Ma,Wong,Hannestad1,Hannestad2,Springel,Viel}. Here, we present a new
calculation with updated parameters and a more realistic halo model.
We also examine weak gravitational lensing as a method for an
astrophysical detection of the cosmic relic neutrinos. We find that
weak gravitational lensing all-sky surveys, such as the planned EUCLID
mission, may detect the presence of the neutrino perturbation in the
average mass density profile of clusters of galaxies, although systematic
uncertainties related to the impact of baryons on the redistribution of
the total mass profile are likely to be severe.

  In Section 2 we describe our method for computing the relic neutrino
clustering within dark matter haloes, and the results are shown in
Section 3. Section 4 discusses how the effect of relic neutrino
clustering within dark matter haloes can be detected by
weak lensing.

  We use the $\Lambda$CDM flat model with $\Omega_M=0.27$ and Hubble constant
$H_0=70 ~ {\rm km \, s}^{-1}\, {\rm Mpc}^{-1}$ throughout the paper,
with a power spectrum normalization $\sigma_8=0.9$ and primordial slope
$n_s=0.96$.

\section{Numerical method}

 This section describes the method we use to compute the neutrino
density profile in a spherical model of the dark matter halo. 

\subsection{Mass density profile}

  In this study we consider neutrinos as test particles moving in a
gravitational potential determined from a spherical model of
the distribution of the cold dark matter, which dominates the total mass
in clusters. Our model adopts the numerical fits that have been obtained
from cosmological numerical simulations of the formation of halos from
cold dark matter. We calculate a density profile including the inner
virialized region and the outer infall region of a halo by smoothly
joining two different pieces. The first piece is the NFW profile
\cite{NFW}, valid inside the virial radius. The second piece is
obtained starting from the average initial density perturbation around
a halo in a Gaussian random field, and evolving it in the non-linear
regime by assuming spherical gravitational collapse without shell-crossing
\cite{bbks,EL}. The two pieces are joined together at an assumed epoch of
observation and at a certain radius, which is determined by requiring
continuity in the density profile (the
derivative of the density profile is allowed to be discontinuous at
the junction point).

 The NFW profile has two parameters, the halo mass and its concentration
parameter, and is given by
\begin{equation}
\rho_{_{NFW}}(r)=\frac{\rho_s}{(r/r_s)(1+r/r_s)^2} ~ ,
\end{equation}
where the concentration parameter is $c=r_v/r_s$, and the virial radius
$r_v$ is obtained from the halo mass as
\begin{equation}
M=\frac{4\pi}{3} \rho_c\Delta_c r_v^3 ~ .
\label{rvirial}
\end{equation}
Here, $\rho_c$ is the critical density of the universe at redshift z,
and $\Delta_c$ is the halo mean density within the virial radius in
units of the critical density, which
for a flat universe with a cosmological constant is given by
\cite{Bryan/Norman}:
\begin{eqnarray}
\Delta_c&=&18\pi^2+82x-39x^2 ~ , \\
\nonumber\\
x&=&\Omega(z)-1 ~ , \\
\nonumber\\
\Omega(z)&=&\frac{\Omega_m(1+z)^3}{\Omega_m(1+z)^3+\Omega_\Lambda} ~ .
\end{eqnarray}
The NFW profile is a fit to the density profile of the halo obtained in
numerical simulations for the virialized region. Outside this region, we
use instead a density profile obtained from the mean mass distribution
around any mass concentration, in a Gaussian field with power spectrum
$P(k)$ \cite{EL}. Let the rms
mass fluctuation within a sphere of radius $r$ be $\sigma_M(r)$. The
average linear overdensity $\overline\delta_2 = \nu_2\sigma_M(r_2)$
within a radius $r_2$, under the condition that the mean linear
overdensity within the smaller radius $r_1$ is equal to
$\overline\delta_1=\nu_1 \sigma_M(r_1)$, can be calculated as
$\nu_2 = \gamma_{12}\nu_1$, where
\begin{equation}
\gamma_{12}=\frac{9}{2\pi^2 \sigma_M(r_1) \sigma_M(r_2)}
 \int_0^\infty dk~P(k)~\bigg[\frac{j_1(kr_1)}{r_1}\bigg]\bigg[\frac{j_1(kr_2)}{r_2}\bigg] ~,
\label{gamma12}
\end{equation}
where $j_1$ is the spherical Bessel function. We use this equation to
obtain the average linear overdensity profile around a halo.

  The outer halo density profile beyond a certain radius $r_{f0}$,
which is to be determined by a matching condition that is specified
below, is then calculated as follows:
we start with a guessed value of $r_{f0}$ with a mean interior
overdensity $\overline\delta_{f0}$ in the NFW profile. The corresponding
initial radius $r_{i0}$ is obtained from $(1+\overline{\delta}_{f0}) =
(r_{i0}/r_{f0})^3$.
Assuming the spherical collapse model with no
shell-crossing (i.e., a constant interior mass), we
calculate the required extrapolated linear overdensity
$\overline{\delta}_{i0}(r_{i0})$ to produce the final overdensity
$\overline{\delta}_{f0}(r_{f0})$. We then evaluate
the linear mean overdensity $\overline{\delta}_i(r_i)$ at
any radius $r_i > r_{i0}$ with equation (\ref{gamma12}),
using the power spectrum of Eisenstein \& Hu \cite{EisensteinHu} with
the parameters $\Omega_m=0.27$, $\Omega_\Lambda=0.73$, $\sigma_8=0.9$,
$n_s=0.96$ and $h=0.7$. Finally, using again the spherical collapse
model, we compute the final radius $r_f$ corresponding to each initial
radius $r_i$ and its linear overdensity $\overline{\delta}_i(r_i)$.
The non-linear density profile is
\begin{equation}
\rho (r_f) = \rho_m \left( \frac{r_i}{r_f} \right)^2 \frac{dr_i}{dr_f} ~ ,
\end{equation}
where the mean density of the universe is $\rho_m = \Omega(z)\rho_c$.

\begin{figure}
\begin{center}
\includegraphics[width=8.5cm]{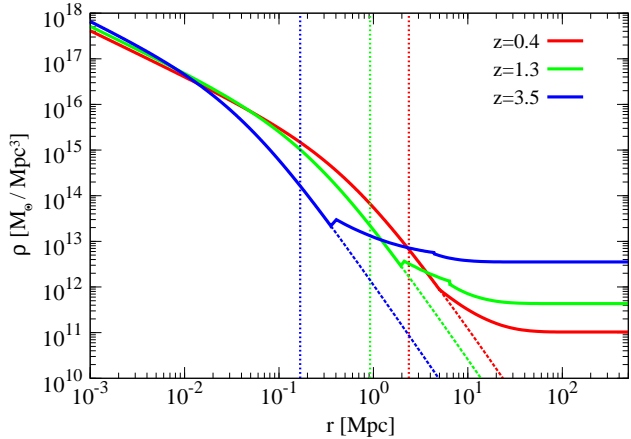}
\caption{Red solid line shows the adopted mean dark matter halo density
profile at redshift $z=0.4$ for $M_{\Delta}=10^{15} h^{-1}\msun$.
The inner profile is the NFW
model, matched with the outer profile computed from the average
spherical perturbation around a halo evolved according to spherical
collapse. The dashed line shows the extrapolation of the NFW model
beyond the matching point. The green and blue lines show the density
profile of the most massive progenitor at redshifts $z=1.3$ and $z=3.5$,
respectively, with masses ($M_{\Delta}$) $2.1\times10^{14} h^{-1}\msun$ and $8.5\times10^{12} h^{-1}\msun$. The density profiles of these
halos are not continuous at the matching point, and their outer profile
is determined by mass conservation as required for assembling the halo
at $z=0.4$. The vertical dotted lines indicate the
position of the virial radius for the three halos. The matching point
chosen for continuity generally occurs around twice the virial radius.}
\label{perfilEL}
\end{center}
\end{figure}

We choose the radius $r_{f0}$ at which the inner NFW and the outer
infall density profile are matched by requiring continuity of the mass
density profile. The matching point that results from this continuity
requirement at a specified redshift is located in all our calculations
between 1.5 and 3 times the virial radius.
In Figure \ref{perfilEL}, the density profile generated using
this procedure is plotted for a dark matter halo of mass 
$M=10^{15} h^{-1}\msun$ at $z=0.4$ (red solid line). The 
dashed line shows the extrapolated NFW profile beyond the
matching point.
At large radius, the mean density profile obviously
approaches the mean density of the universe. The figure also shows the
profiles of other halos at a higher redshift with the average mass of
the most massive progenitor of the halo at $z=0.4$, as discussed
below.
The vertical dotted lines indicate the position of the
virial radius of each halo.

\subsection{Dark matter halo evolution}

  Modeling the orbits of relic neutrinos in a cluster halo requires the
gravitational potential of the halo to be specified as a function of
time. To obtain a realistic model for a typical halo, we use the results
of \cite{Zhao} to obtain a mass of the halo as a function of time over
its entire history of accretion. Obviously, there is a large dispersion
in the accretion history of a halo and therefore in the evolution of its
potential well, but we take an average history for the most massive
progenitor as a typical case to calculate the orbits of the neutrinos in
our spherical model.

  We use the empirical formula of \cite{Zhao}, obtained from a numerical
fit to the results of N-body cosmological simulations, to calculate the
mass and concentration parameter of the most massive progenitor of our
halo of mass $M_f$ at the final redshift $z_f$, for each earlier epoch
at redshift $z>z_f$. This is done for 100 values of the redshift $z$,
distributed logarithmically between $z=z_f$ and $z=10$.

  As this work was being carried out, we initially computed the density
profile of the progenitor halo of mass $M_h(z)$ at each redshift $z$
with the same method as for the final halo at redshift $z_f$, choosing a
matching radius and requiring continuity with the average external
density profile. However, this method does not conserve the total mass
because it does not take into account the requirement that the
mass $M_f$ that is assembled into the final halo at $z_f$ must be
present in the external region around the progenitor halos within the
shell that will finally collapse onto the halo at redshift $z_f$. In
other words, the density profile around a progenitor halo of mass
$M_h(z)$ is not equal to the average one as obtained from equation
(\ref{gamma12}), but is modified by the condition
that a halo of mass $M_f$ must be assembled at redshift $z_f$.
Therefore, the density profile of the progenitor halo is computed by
fixing the matching point to the same fixed multiple of the virial
radius $r_v$ as for the final halo at $z_f$, $r_{f0}(z)= r_{f0}(z_f)
\times r_v(z)/r_v(z_f)$, and tracing back in time the
position of each spherical shell around the halo. At each step in
redshift (backwards in time), the progenitor halo decreases its mass
within $r_{f0}(z)$ by an amount $\delta M$, and a new spherical
shell is added with mass $\delta M$ with a radius equal to $r_{f0}(z)$.
All the spherical shells are traced back in time using the
spherical collapse model with no shell-crossing. This results in the
density profiles shown in Figure \ref{perfilEL} for two examples of the
progenitor halos, at $z=1.3$ (with mass $M=2.1\times10^{14}\,
h^{-1}\msun$) and at $z=3.5$ (with mass $M=8.5\times10^{12}\,
h^{-1}\msun$). The density profile is no longer continuous at the
matching point, but this does not cause any problem.

  We have found that correctly computing the evolution of the conserved
external mass distribution around the halo of a cluster is important: if
one uses instead the mean density profile around a halo progenitor, the
final result for the neutrino density can be underestimated by more than
a factor of two.

  The evolving potential of the halo is computed by interpolation from
the mass profiles calculated at 100 values of the redshift $z$, as the
orbits of test particles representing the neutrinos are integrated.

\subsection{Neutrino orbits}

  The initial phase space distribution of neutrinos is determined by
their state of thermal equilibrium reached in the early universe with
the primordial plasma, i.e., the Fermi-Dirac distribution for highly
relativistic particles,
\begin{equation}
f(p)\, dp=\frac{8\pi}{(2\pi \hbar)^3}\frac{p^2dp}{e^{p/T}+1} ~ .
\label{distribution}
\end{equation} 
The neutrino temperature, evolving as $T=T_0(1+z)$, is related to
the photon temperature $T_{\gamma 0}$ as $T_0=(4/11)^{1/3}T_{\gamma 0}
 \approx 1.9$ K.

  The orbits followed by neutrinos in our time-dependent spherical
potential depend on three orbital parameters: the initial radius and
momentum, and the angular momentum of the neutrino. To compute the
neutrino density profile, this three-parameter space of neutrino
orbits needs to be sampled densely enough to compute their average
spatial distribution as a function of time. For this purpose, we
divide the initial radius, momentum and angular momentum into several
bins, and compute a neutrino orbit for each binned value of the three
variables, starting the orbits at $z_i=10$. This three-dimensional
phase-space grid is constructed taking particular care to resolve
the particles reaching close to the center of the halo, which are at
small initial radius or small angular momentum.

  The grid is constructed using 10000 bins in radius from 0 to
$r_{max}$, distributed as $r_i=r_{max}(i/10000)^2$,
where $r_{max}$ is large enough to ensure convergence of the final
neutrino density profile out to a final radius of at least 30 Mpc.
Momentum bins are similarly set by $p_j=p_{max}(j/500)^2$, with 500
bins, where $p_{max}=0.005(1+z_i)$ eV, sufficient to sample particles
out to the largest momenta making any significant contribution. Finally,
the angular momentum is sampled from 0 to $L_{max} = r_ip_j$ using 200
bins distributed as
\begin{equation}
\left \{
\begin{array}{rl}
\theta_k=\left(\frac{\pi}{2}\right)\left(\frac{k}{100}\right)^{\alpha}
& (k \leqslant 100) ~ , \\
\\
\theta_k=\pi-\left(\frac{\pi}{2}\right)\left(\frac{k-100}{100}\right)^{\alpha}
 & (k>100) ~ ,
\end{array} \right.
\label{NFWcutoff}
\end{equation}
where $\theta_k$ is the angle subtended between the initial momentum and
radius of the particle. Here, $\alpha$ is a parameter to control the
sampling of particles with low angular momentum, which are responsible for
the shape of the density profile in the inner parts. Typically, it
ranges between 1.5 to 5 depending on both neutrino and dark matter halo
mass. At each three-dimensional bin, neutrino orbits are computed by solving
the equation
\begin{equation}
\frac{d^2r}{dt^2}-\frac{L^2}{r^3}=-\frac{\partial \phi (r,t)}{\partial r} ~ ,
\label{eqnmov}
\end{equation}
where $L$ is the conserved angular momentum per unit of mass and $\phi$
is the time-dependent Newtonian gravitational potential, computed from
the dark matter density profile specified in Section 2.1. The contribution of
each neutrino particle to the final neutrino density profile as a function of
time is counted as a spherical shell of radius $r(t)$ and mass
proportional to the weight of the bin at radius $r_i$, momentum
$p_j$ and angle $\theta_k$ in the phase space distribution,
\begin{equation}
m_p^{i,j,k} \propto \int _{r{_i}}^{r_{i+1}} r^2\,dr
\int_{p_j}^{p_{j+1}} \frac{p^2dp}{e^{p/T_\nu(z)}+1}\times
\| \cos\theta_{k+1}-\cos\theta_k \| ~ . 
\end{equation}
The final neutrino density profile is obtained by adding the mass of all interior
shells at any radius and time. Equation (\ref{eqnmov}) is solved for
each particle with a Runge-Kutta fourth-order integrator with variable
stepsize. 

\section{Neutrino density profiles}

We now present the results for the spherical neutrino density profiles. We will discuss four neutrino mass
 schemes: a) three neutrinos with $m=0.3$ eV (labelled 0.3 eV), b) three neutrinos with m=0.15 eV (labelled 0.15 eV), 
 c) two neutrinos with 0.05 eV and one massless neutrino (labelled IH 0.05eV) and d) one neutrino with 0.05 eV and two 
 massless neutrinos (labelled NH 0.05eV). We neglect the mass
squared differences in schemes a and b and the small mass squared
difference in schemes c and d. 
This approximation is justified because the masses of the neutrinos that
we consider to have equal mass differ by less than
1 \% (scheme a) , 5\% (scheme b), and 2\% (scheme c). In scheme c and d,
the neutrinos that we neglect have masses smaller than 0.01 eV, and as
we shall see their contribution to the total neutrino mass profile is
indeed negligible. With this approximation, neutrino density profiles need to be
computed only for masses of 0.3, 0.15 and 0.05 eV.

\begin{figure*}
\begin{center}
\includegraphics[width=7.5cm]{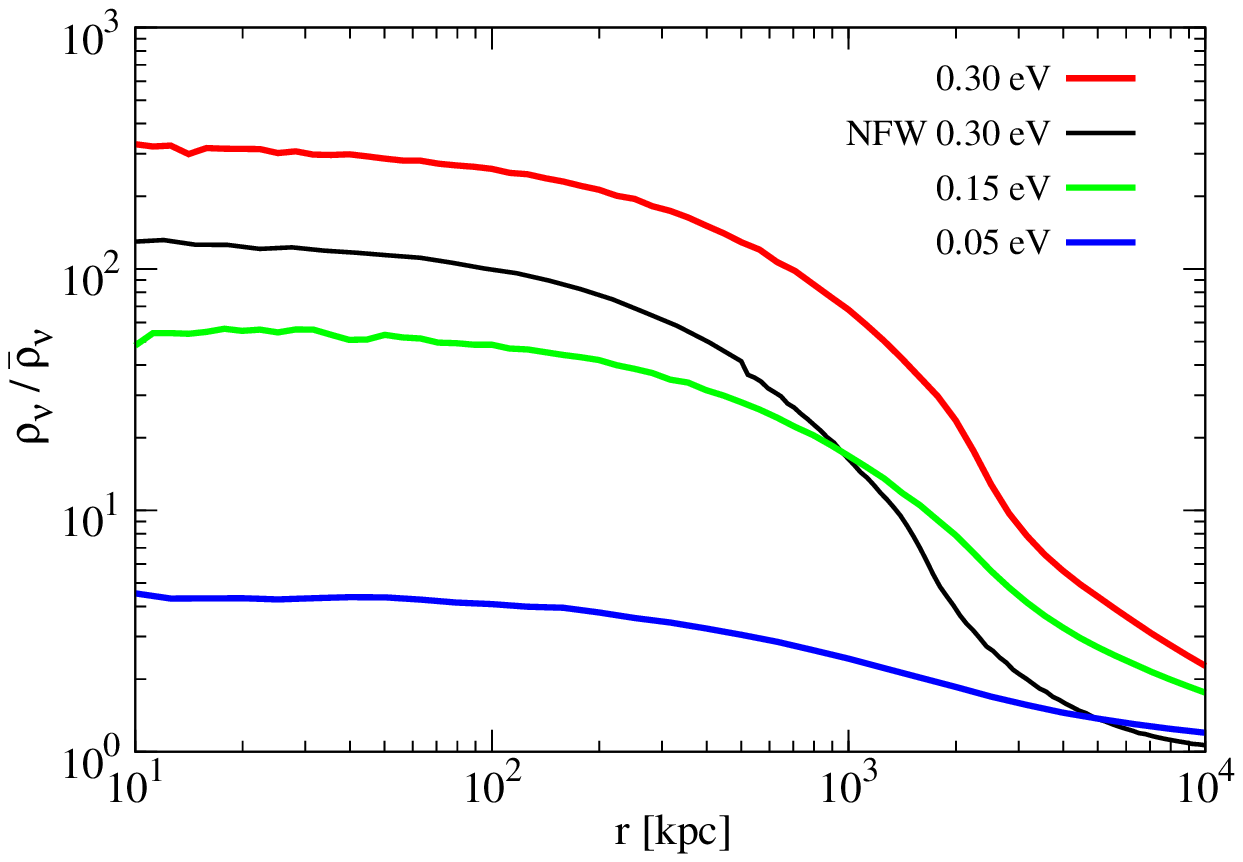}
\includegraphics[width=7.5cm]{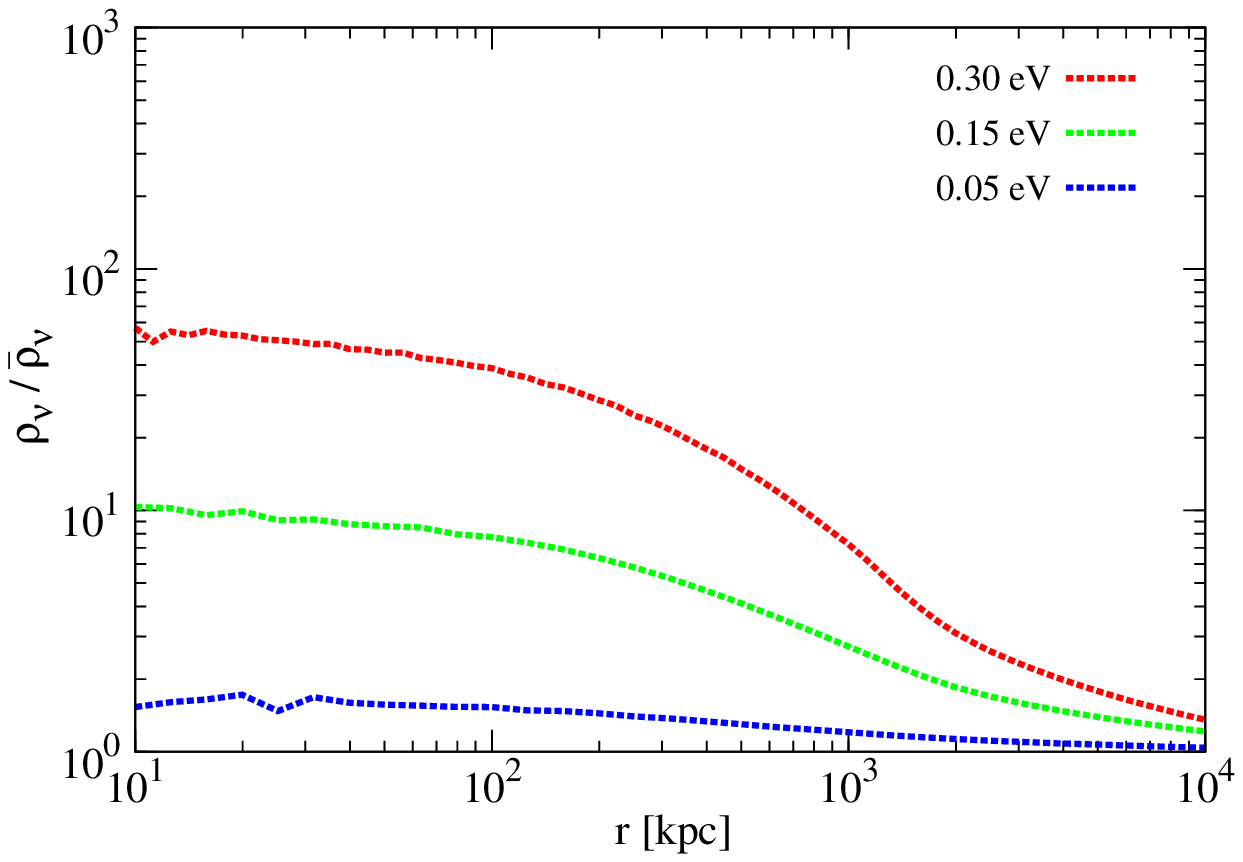}
\caption{Neutrino density profiles at $z=0.4$, shown as the ratio to the
mean neutrino density, for a halo mass of
$10^{15}\, h^{-1} M_{\odot}$ (left panel) and
$10^{14}\, h^{-1} M_{\odot}$ (right panel), and for the
indicated neutrino masses. The case where the NFW profile is used at all
radii (with a suppressed density at large radius, see Figure \ref{perfilEL}) is shown
for one case in the left panel as a black line.}
\label{profile}
\end{center}
\end{figure*}

  Neutrino density profiles are shown at $z=0.4$ in Figure \ref{profile}
in units of the mean cosmic neutrino density, for halos of mass
$10^{15}\, h^{-1} M_{\odot}$ (left) and $10^{14}\, h^{-1} M_\odot$
(right), and for neutrino masses of 0.05, 0.15 and 0.3 eV.
The neutrino overdensity increases with both neutrino mass and halo mass,
as the ratio of the halo velocity dispersion to the neutrino thermal
velocities increases. The size of the core of the neutrino distribution
decreases rapidly with this ratio owing to phase space density
conservation. The random oscillations at small radius are due to
numerical noise arising from the number of particles representing
spherical shells in our simulation.

  The left panel also shows, for a neutrino mass of 0.3 eV, the case of
the NFW mass profile extended over all radii. This results in a reduced
density, as seen in Figure \ref{perfilEL}. The reduction of
the depth of the potential well in this model reduces the neutrino
density.

\begin{figure*}
\begin{center}
\includegraphics[width=8.5cm]{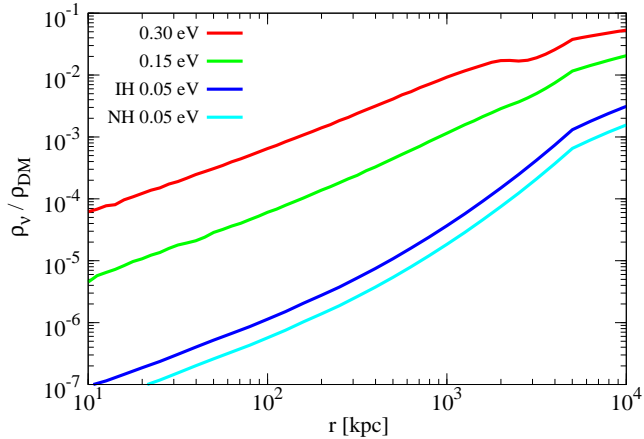}
\caption{The neutrino to dark matter density profiles ratio at
$z=0.4$ for the indicated neutrino masses, in a halo of mass of
$10^{15}\, h^{-1} M_{\odot}$.}
\label{caustics}
\end{center}
\end{figure*}

  In Figure \ref{caustics}, the ratio of the neutrino to the dark matter
mass density profile, $\rho_\nu/\rho_{DM}$, is plotted for a halo of
mass $10^{15}\, h^{-1} M_\odot$ at $z=0.4$, for neutrino masses of 0.3,
0.15 and 0.05 eV. Schemes a and b are used for the two heavier masses
(i.e., the red and green curves show the density computed for one
neutrino family multiplied by 3), and schemes c and for the lighter
mass (the cyan curve is for one neutrino family, and the blue curve is
for two). A change of slope occurs at a radius close to 5 Mpc, due to
the change of slope in the
mass density profile at the matching point between the NFW and the
outer infall model of the average density perturbation. Near a radius
of 2 Mpc, a feature is present that is particularly strong for the
largest neutrino mass and becomes weak as the neutrino mass is
decreased. This is the result of a caustic, a special feature of
spherical collapse. For perfectly cold particles, a true caustic (where
the density becomes formally infinite) would appear at this radius, at
which the single infalling stream of particles outside the caustic
changes to a superposition of three streams inside the caustic, owing to
the particles that are turning around in their first orbit after going
through the halo center. The
caustic is increasingly smoothed out for neutrinos as their primordial
velocity dispersion increases (i.e., the neutrino mass decreases), or
as the halo mass decreases. In practice, this caustic feature is
present only in a spherically symmetric system. Real clusters collapsing
from random initial density perturbations have caustics that are highly
irregular and occur at variable radii, influenced by their internal
substructure and non-sphericity, and which are largely washed out when
averaging over many clusters (see, e.g.,  \cite{Fillmore84,Vogel}).
Note also that a caustic should of course also be present in the Cold
Dark Matter in a spherical model, which we are not taking into account
here because we are using a simple analytic model for the mass profile.
The Cold Dark Matter would have its caustic
washed out only by the effects of substructure and non-sphericity,
while the neutrino caustic is further washed out by the initial
thermal velocities.

Comparing our calculations with previously published results, we find
that we reproduce the results by \cite{Wong} when using their dark
matter halo (NFW) density profile and halo evolution model, but we do
not reproduce those of \cite{Ma} (see \cite{Wong} for a discussion of
this difference). As we have shown, the NFW halo profile extrapolated to
large radius that is used by \cite{Wong} underestimates the neutrino
contribution to the profile at large radius. Our model also improves
that of \cite{Wong} on the cluster evolution, by including the mean
redshift dependence of the halo progenitor mass and concentration
parameter, instead of a constant halo mass during the accretion history
used in \cite{Wong}, and by computing also the mean spherical evolution
of the density profile external to the halo.

\begin{figure*}
\begin{center}
\includegraphics[width=7.5cm]{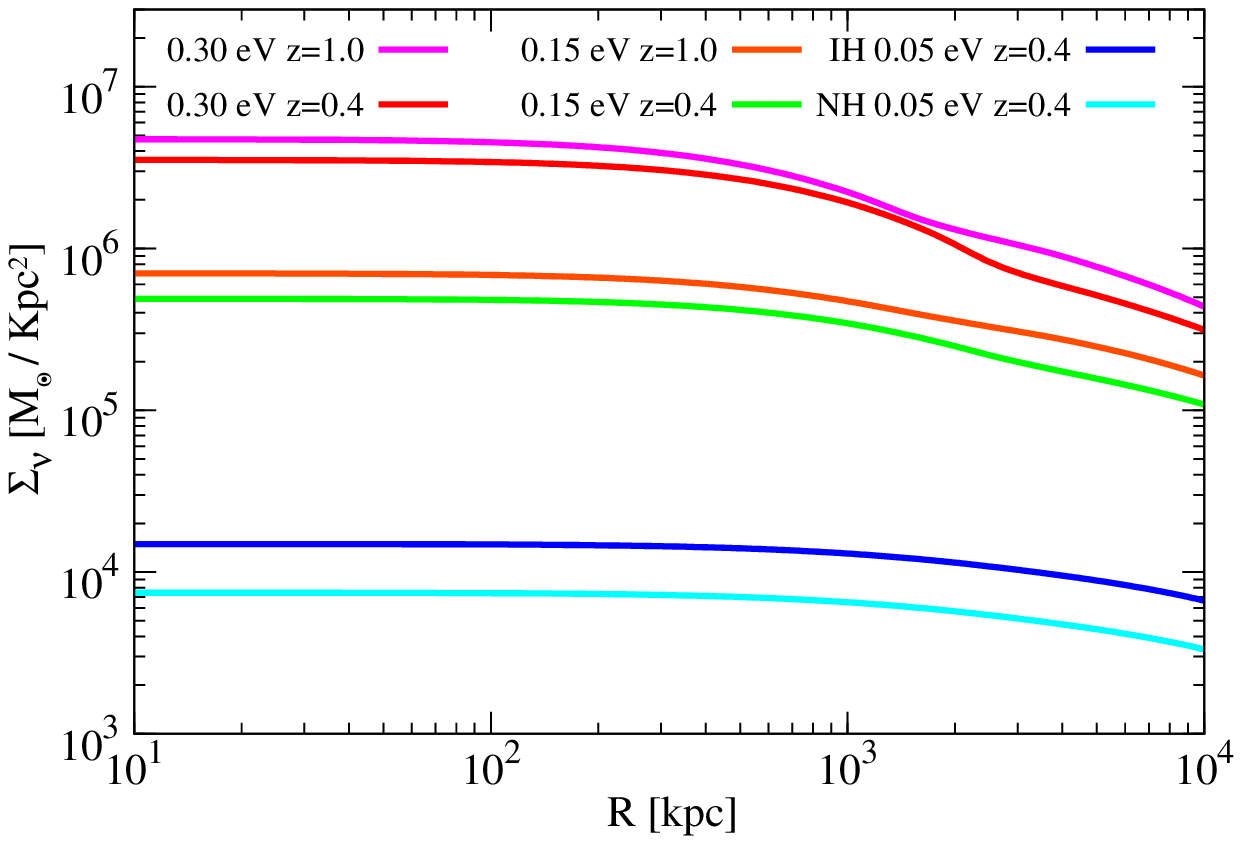}
\includegraphics[width=7.5cm]{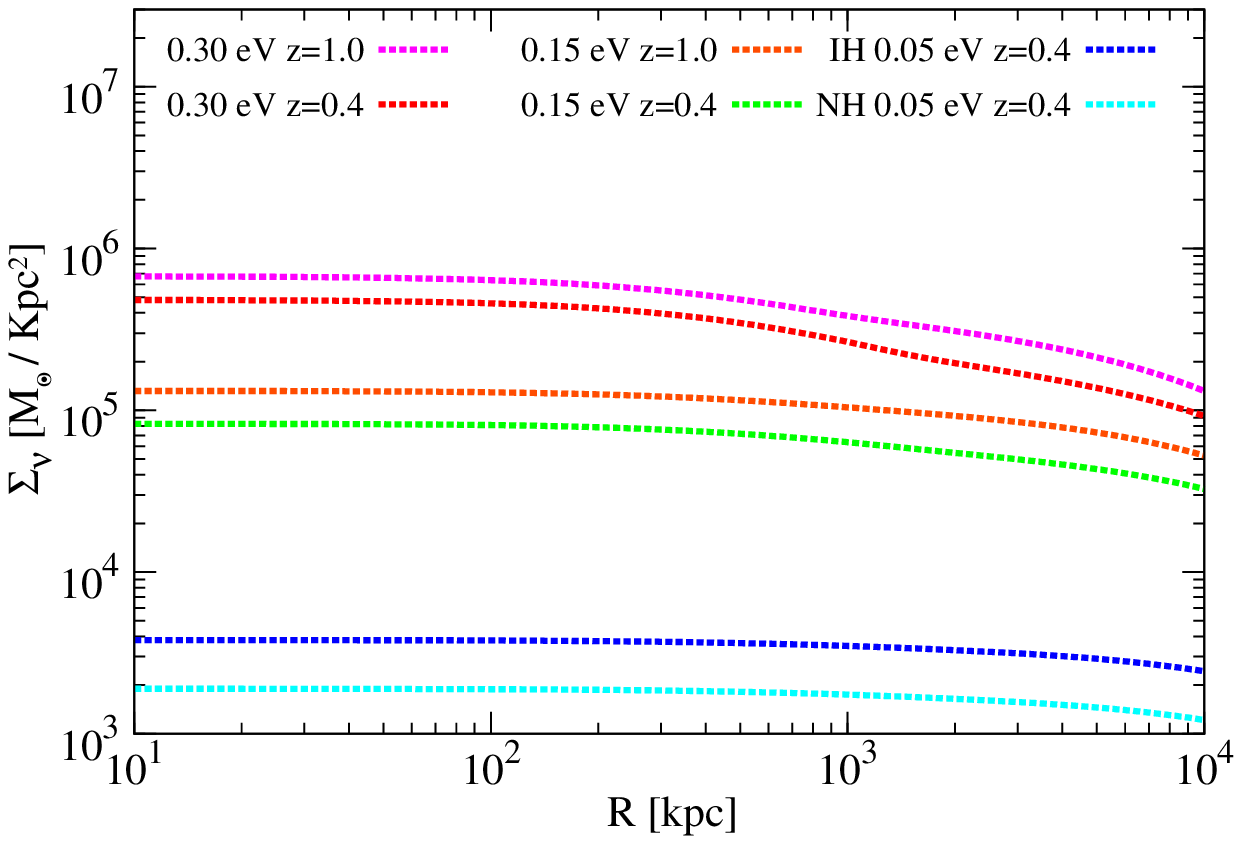}
\caption{Neutrino surface over-density profiles for neutrinos with the
indicated masses, at $z=0.4$ and $z=1$. Left panel is for halo mass
$10^{15}\, h^{-1} M_\odot$, and right panel for $10^{14}\,
h^{-1} M_\odot$.}
\label{surface_density}
\end{center}
\end{figure*}

  Finally, in Figure \ref{surface_density} we plot the neutrino surface
overdensity, which is important for our lensing calculations in the next
section,
\begin{equation}
\Sigma_\nu(R)=\int_{-\infty}^{\infty} \big[\rho_\nu(r) -
\overline{\rho}_\nu\big]\,  dx ~ ,
\end{equation}
where $r^2=x^2+R^2$, $R$ is the projected radius on the sky and $x$ is
the dummy variable for integration along the line-of-sight, and
$\overline{\rho}_\nu$ is the mean neutrino
density, for the same cases of neutrino mass, and at redshifts $z=0.4$
and $z=1$. The left panel is for a halo mass $10^{15}\, h^{-1} M_\odot$,
and the right panel for $10^{14}\, h^{-1} M_\odot$. The same main
effect is clearly seen as previously: the core radius of the neutrinos
is reduced, and their central overdensity increases, as the halo mass or
neutrino mass increases.

\section{Neutrino detection with weak lensing}

  We now consider the possibility of detecting the perturbation caused
by neutrinos on the radial density profile of a cluster using weak
gravitational lensing. In this section we consider the idealized case
where weak lensing can be measured for a large number of clusters
with a perfectly known selection function,
with statistical errors declining as the square root of the
number of clusters.

 We summarize first the basic concepts of weak gravitational lensing.
The distortion of images behind an extended gravitational lens is
determined by the surface density of the lens at every point in
projection on the sky, $\Sigma$.
The convergence is $\kappa = \Sigma/\Sigma_{crit}$,
where $\Sigma_{crit}$ is the critical surface density, which depends
on the angular diameter distances to the lens ($D_l$), to the source
($D_s$), and from the lens to the source ($D_{ls}$):
\begin{equation}
\Sigma_{crit}=\frac{c^2D_s}{4\pi GD_{ls}D_l} ~ .
\end{equation}
In general, a spherical
source acquires an elliptical shape after being lensed, with axis
ratio $(1-\kappa-\gamma)/(1-\kappa+\gamma)$, where $\gamma$ is the
shear (for reviews see, e.g., \cite{Blandford,Schneider}). In a spherical lens, the shear is given by
\begin{equation}
 \gamma(R) = \bar\kappa(R) - \kappa(R) ~,
\label{eqgam}
\end{equation}
where $\bar\kappa(R)$ is the average convergence within a projected
radius $R$,
\begin{equation}
 \bar\kappa(R) = {2\over R^2} \int_0^R dR'\, R'\, \kappa(R') ~.
\label{eqbk}
\end{equation}
The weak lensing limit is the case when $\kappa \ll 1$ and
$\gamma \ll 1$, in which case the ellipticity acquired by the source
is $\epsilon \simeq 2\gamma$.

 For an arbitrary mass distribution without spherical symmetry, we can
choose any center we may wish and consider the values of the convergence
and shear averaged on circles of radius $R$ around the chosen center.
The averaged quantity $\bar\kappa(R)$ is also obtained by
averaging the convergence within a radius $R$. Equation (\ref{eqgam})
is then just as valid for an arbitrary mass distribution, provided that
we define $\kappa(R)$ and $\gamma(R)$ by averaging over circles of
radius $R$ (or, in other words, circularly rotating the lens around the
chosen center and averaging over all possible angles of rotation).

  The quantity $\gamma(R)$ is the one we can directly measure
from the shapes of the lensed galaxies, and the density profile of the
cluster lens can be reconstructed by the use of inversion methods
\cite{lensing}.
A very useful particular case is obtained by considering the integral
\begin{equation}
\int_{R_1}^{R_2} {dR\over R} 2 \gamma(R) =
\bar\kappa(R_1) - \bar\kappa(R_2) \equiv C_{12} ~.
\end{equation}
This equality is easily verified from equations (\ref{eqgam}) and
(\ref{eqbk}). Hence, we can measure {\it differences} in the projected
mass at two different radii $R_1$ and $R_2$, from the directly
observable shear in the annulus between the two radii.

 The shear cannot be measured exactly because the sources have random
ellipticities with dispersion $\sigma_e$. If the sources have a number
density $n$ (considering them to be all at the same redshift for
simplicity), the error in the measurement of the average $\gamma$ in
an annulus of radius $R$ and width $\Delta R$ is
\begin{equation}
 \sigma_{\gamma} = {\sigma_e\over 2} (2\pi R\, n\, \Delta R)^{-1/2} ~ .
\end{equation}
The error in the quantity $C_{12}$ is then given by
\begin{equation}
 \sigma_{C} = {\sigma_e\over 2\sqrt{\pi n R_1^2} \sqrt{1-R_1^2/R_2^2}} ~ .
\end{equation}

 The mean value of $C_{12}$ averaged over a large sample of clusters
depends on the mass distribution of the clusters and any cosmological
parameters that affect the average halo density profiles. Ideally, if a sample
of clusters
is selected in a perfectly controlled way, one can predict their mean
density profile and the function $C_{12}$. The density profile is
affected by neutrinos, and if all other physical factors and selection
effects influencing the mean density profile are correctly
known and taken into account, the presence of neutrinos may be detected
from the observed shape of the cluster shear profile using weak lensing.

\begin{figure}
\begin{center}
\includegraphics[width=8.5cm]{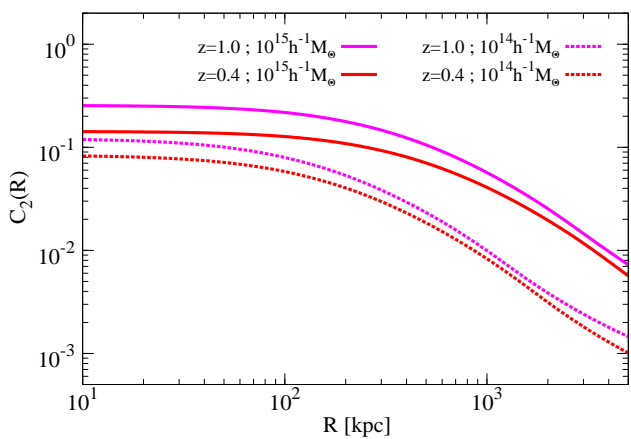}
\caption{The function $C_f(R)$, with $f=2$, for four dark matter halos
with the masses and redshifts indicated.}
\label{kr}
\end{center}
\end{figure}

  As a specific example, we consider the case $R_2= f R_1$, where $f$
is a constant that we fix to $f=2$. In Figure \ref{kr}, the function
$C_f(R) \equiv  \bar\kappa(R) - \bar\kappa(fR)$ is plotted for four
cases, with cluster masses of $10^{15}\, h^{-1}M_\odot$ and
$10^{14}\, h^{-1}M_\odot$ at $z=0.4$, and masses
$10^{15} h^{-1} M_\odot$ and $10^{14} \, h^{-1} M_\odot$ at $z=1$.
The sources are assumed to lie all at $z_s=1.5$.

\begin{figure*}
\begin{center}
\includegraphics[width=7.5cm]{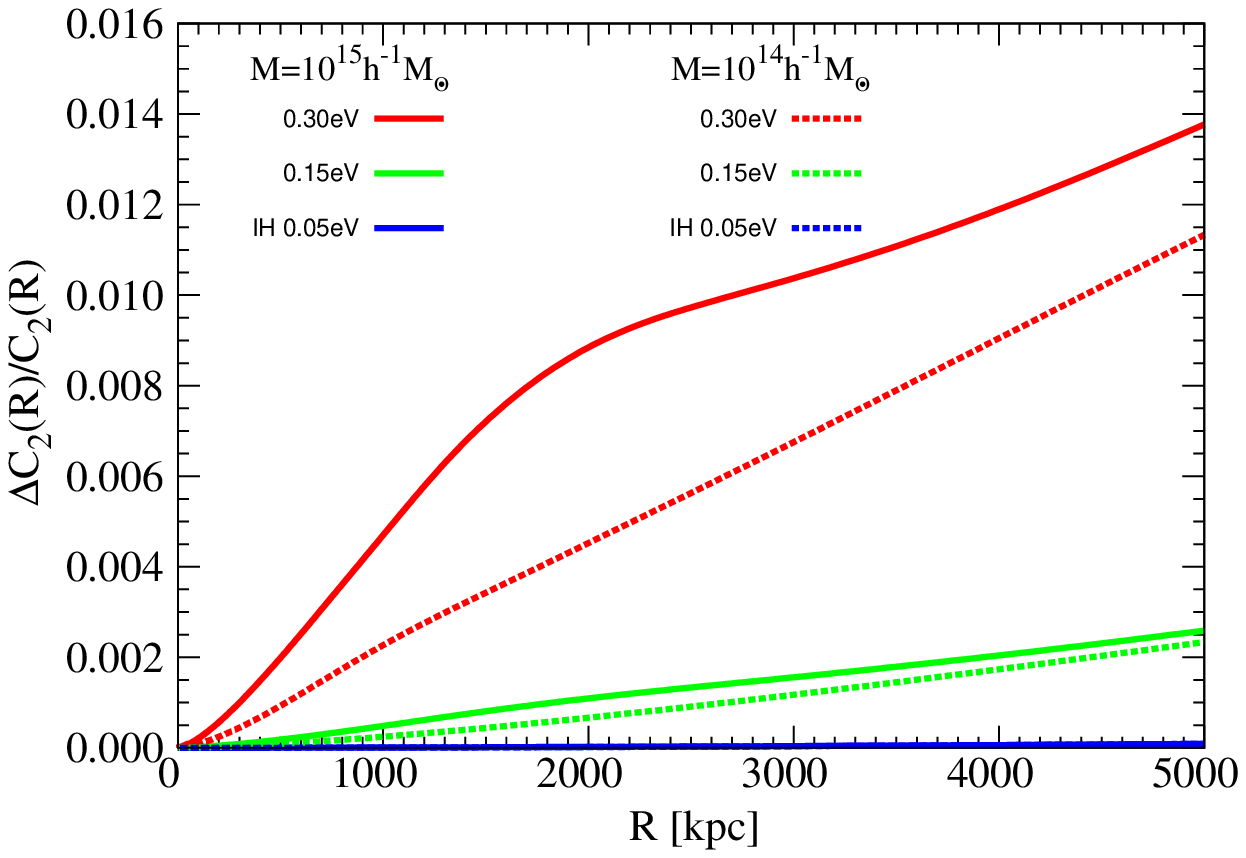}
\includegraphics[width=7.5cm]{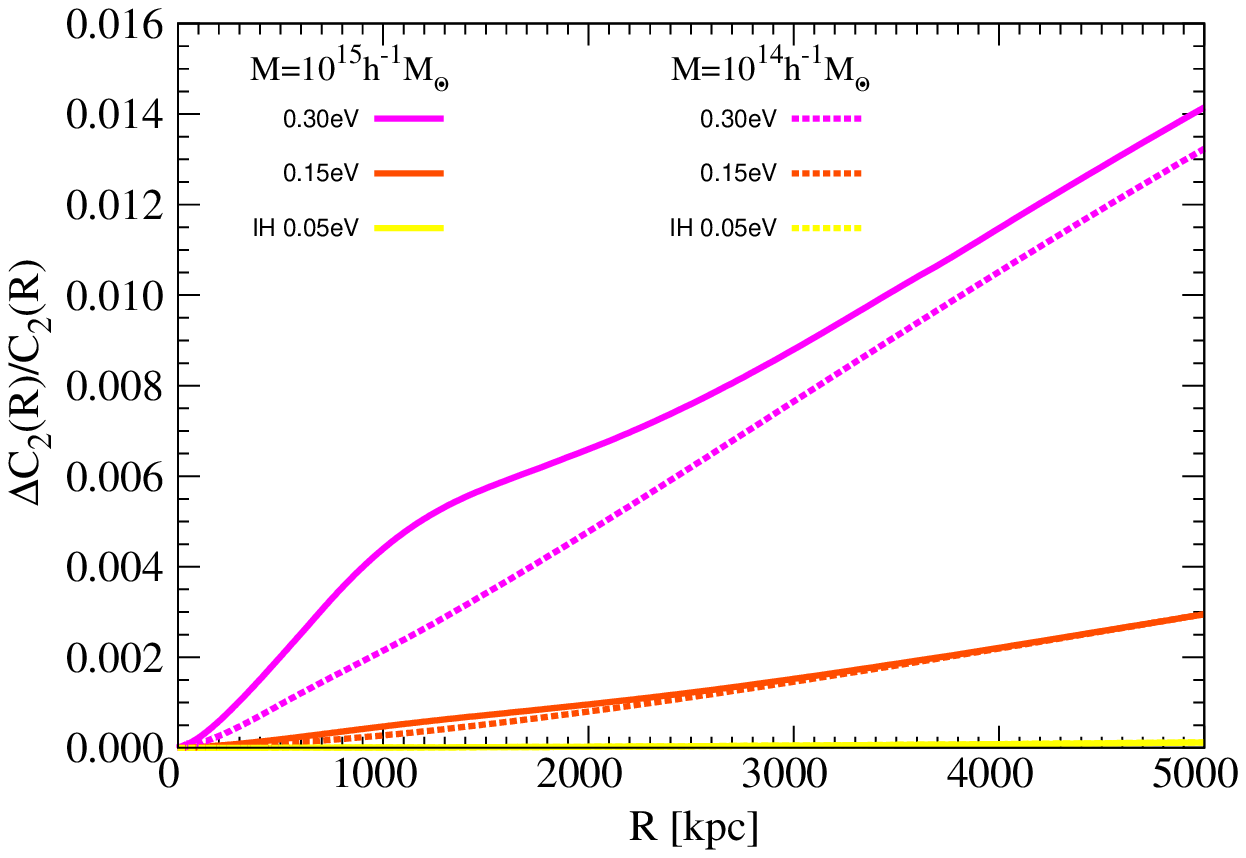}
\caption{Fractional neutrino perturbation on the weak lensing profile,
$\Delta C_f/C_f$, versus radius, for the cluster masses and neutrino
masses indicated. The left panel is for clusters at $z=0.4$, and the
right panel at $z=1$. }
\label{deltaC}
\end{center}
\end{figure*}

\begin{figure*}
\begin{center}
\includegraphics[width=7.5cm]{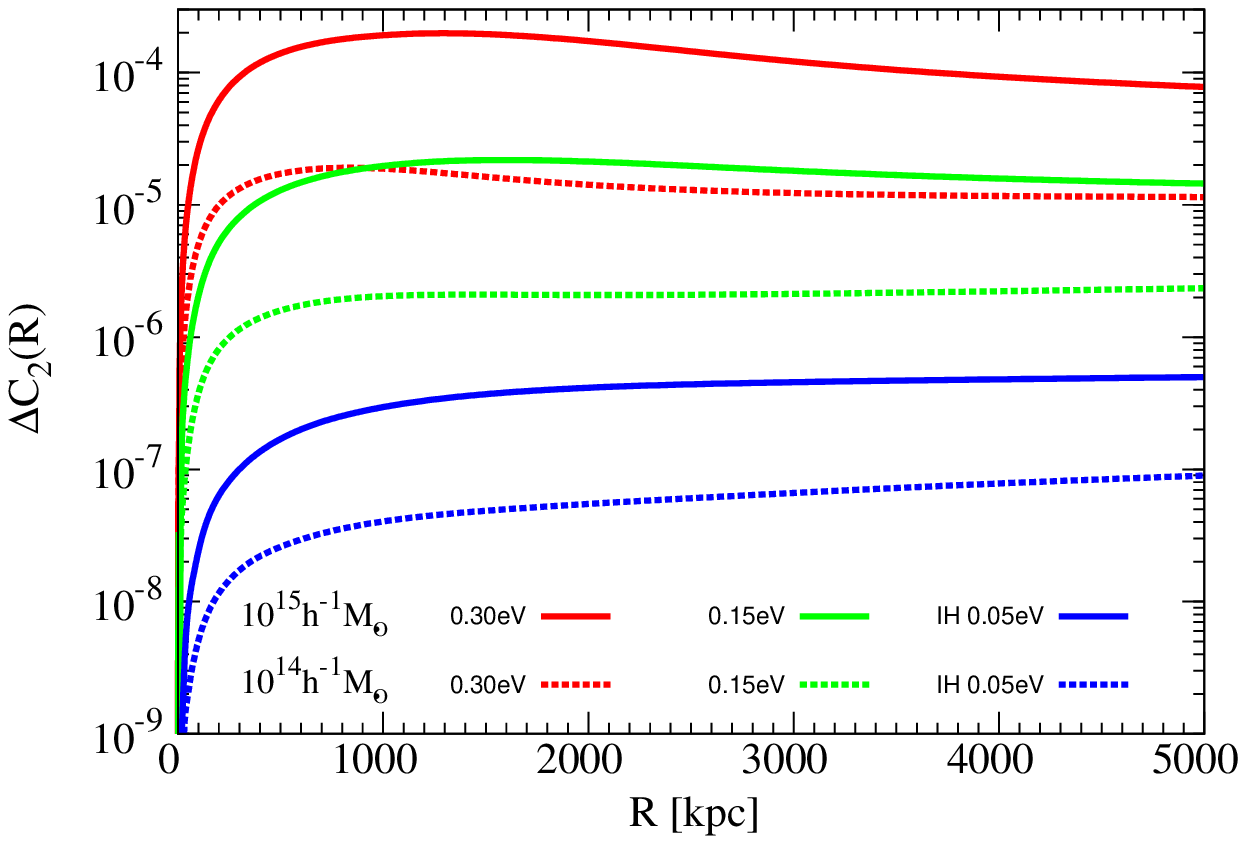}
\includegraphics[width=7.5cm]{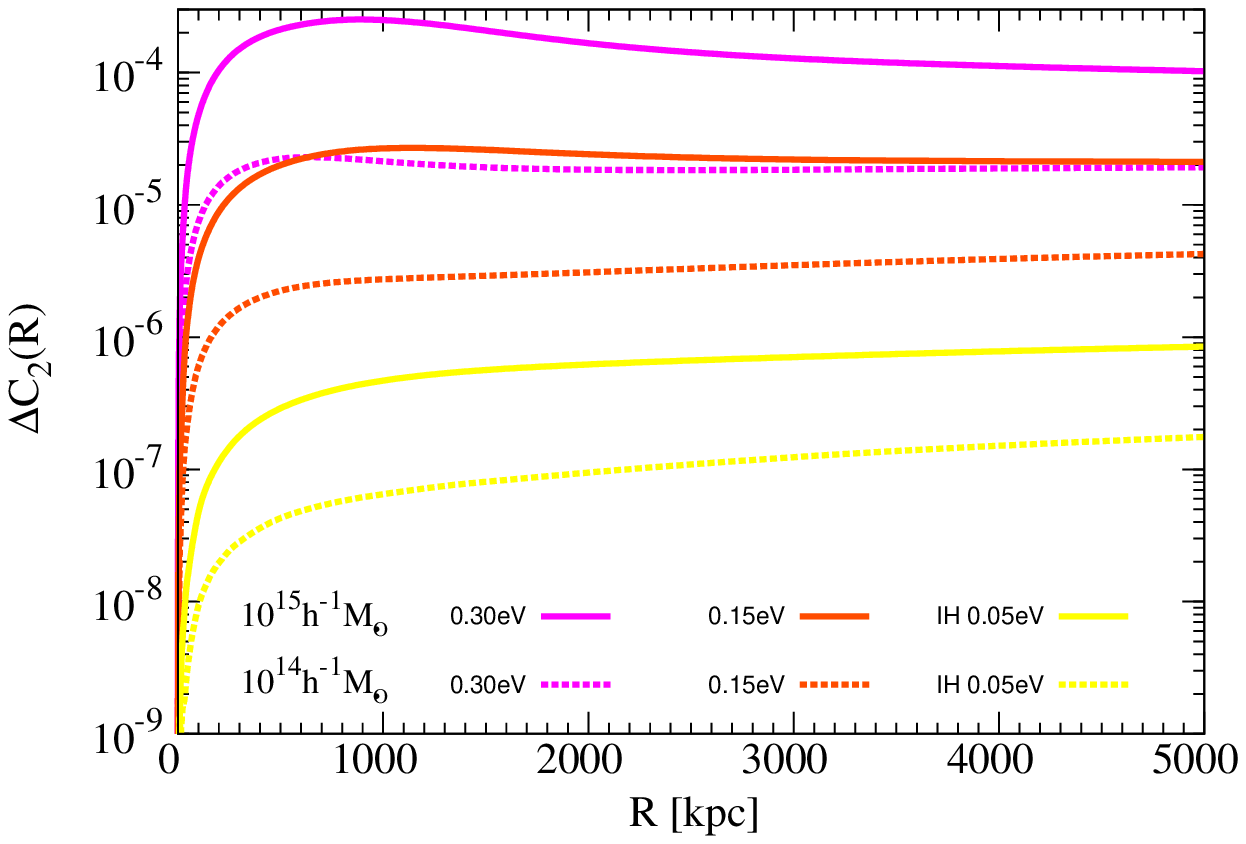}
\caption{Neutrino perturbation on the weak lensing profile,
$\Delta C_f$, versus radius, for the same cluster and neutrino masses as
in Figure \ref{deltaC}. Left panel is for halos at $z=0.4$ and right panel at $z=1$.}
\label{absC}
\end{center}
\end{figure*}

  The effect of neutrinos is to modify the observable function
$C_{f}(R)$ by a fractional amount $\Delta C_f/C_f$, where $\Delta C_f$
is calculated for the neutrino density profile in the same way as $C_f$
for the total mass profile. This ratio is plotted
in Figure \ref{deltaC} for various neutrino masses and for two
different dark matter halo masses. The ratio increases with neutrino
mass and grows with radius because the neutrino density profile is extended.
For the cases that are shown, the fractional weak lensing effect of
neutrinos does not change much with halo mass,
although the observable effect, $\Delta C_f$, obviously increases with
halo mass, as shown in Figure \ref{absC}.

The predicted neutrino effect is very small, and it might only be
observable as a perturbation in the mean cluster shear profile by averaging
over many clusters. The requirement for detecting the neutrino effect
can be estimated by considering the cluster in Figure \ref{kr} with a
halo mass $10^{15}\, h^{-1}M_\odot$ at $z=0.4$, and sources
with number density $n = 30 \, {\rm arcmin}^{-2}$ located at $z=1.5$
and at the radius $r_1= 2$ Mpc (corresponding to 6 arc minutes).
The 1-$\sigma$ error on $C_f$ is $\sigma_C = 0.002$ if we use
$\sigma_e=0.2$, while its value is $C_f(r_1)\simeq 0.01$. Therefore, we may
reach an accuracy of 20\% on the measurement of $C_f$ with a single
cluster. To be able to measure the difference between different neutrino
models of $\Delta C_f / C_f < 0.01$, as expected from Figure \ref{deltaC} for
a neutrino mass of $0.3$ eV, one would need to average the
measurement of the shear over 10000 clusters to obtain a 5-$\sigma$
result.

  This is approximately the number of massive clusters that might be
observed in an all-sky weak lensing survey of sufficient depth.
Therefore, the measurement of the neutrino perturbation on the mean
density profile of clusters is extremely difficult. Apart from the
need to observe a very large number of clusters to reduce the
statistical error, systematic uncertainties would in practice be
even more difficult to resolve. The theoretical prediction for the
precise density profile in the absence of neutrinos needs to be
sufficiently reliable, but this profile is affected by several variables
that may be hard to control: the precise selection function of clusters
of different masses and different spatial orientations and projection
effects would need to be accurately modeled using numerical simulations
of structure formation, and the contribution from baryons would be
subject to uncertainties related to radiative cooling, galactic winds,
and generally the way that galaxy formation may alter the mass
distribution. The detection of the gravitational effect of neutrinos
from lensing seems therefore a very difficult challenge.

  The calculation presented in this paper should be considered only
as an illustrative case. In practice, a better approach for attempting
to measure the clustering effects of neutrinos may be to examine
directly the power spectrum and bispectrum of the weak lensing shear
over the whole field, thereby avoiding the issue of selection
effects in a cluster sample. However, this would necessarily average out
the effects of neutrinos in the regions where they are strongest, in
massive clusters of galaxies. The cross-correlation of lensing shear
with massive galaxies or diffuse X-ray emission that are associated with
clusters would also likely be subject to similar uncertainties arising
from the precise selection function.

\section{Conclusion}

  We have presented the clustering of relic neutrinos around spherical
dark matter for various illustrative cases. Neutrinos produce an
extended distribution of mass with a large core determined by their
primordial thermal velocities, which cause a perturbation on the total
density profile. The non-linear collapse of neutrinos in massive
clusters should modify their impact on the overall mass power spectrum
of fluctuations calculated in the linear regime.

  The presence of the neutrino perturbation in the average mass density
profile of clusters of galaxies using weak gravitational lensing would
constitute a remarkable astrophysical detection of the cosmic relic
neutrinos, which cannot be detected by any other known method, except
for their linear contribution to the total matter power spectrum.
However, this measurement is a very difficult one owing to the small
contribution that neutrinos make to the cluster mass even at very
large radius. For a neutrino mass of 0.3 eV, the largest value that is
compatible with current experimental constraints, the lensing shear
profile of a massive cluster is affected by neutrinos roughly at the
level of 1\%. This small signal can only be detected by averaging
lensing measurements over observable clusters in a large fraction of
the sky. Although this observation can be done with
an all-sky weak lensing space mission, such as the planned EUCLID
mission, systematic uncertainties related to the
impact of physical effects such as the distribution of baryons and the
precise cluster selection function would make this detection a
difficult one.

\section*{Acknowledgments}
FV would like to thank Neal Dalal, Urbano Fran\c{c}a, Olga Mena and Manuel Pe\~na for 
discussions and revisions of this paper. The numerical computations  were carried out in the
Sunnyvale cluster at the Canadian Institute for Theoretical Astrophysics (CITA). This work is 
supported in part by the Spanish MICINN grants
FPA-2007-60323, AYA2009-09745, AYA2010-21322-C03-01, the Consolider Ingenio-2010 project
CSD2007-00060 and the Generalitat Valenciana grant PROMETEO/2009/116.

\end{document}